\documentclass[12pt]{article}
\usepackage{graphics}
\begin{document}
\title{Solute Effects on the Helix-Coil Transition} 
\author{Oded Farago and Philip Pincus \\
Materials Research Laboratory, University of California,\\
Santa Barbara, California 93106}
\maketitle
\begin{abstract}
  We discuss the effects of the solvent composition on the helix-coil
  transition of a polypeptide chain. We use a simple model to
  demonstrate that improving the hydrogen-bonding ability of the
  solvent can make the transition less cooperative, without affecting
  the transition temperature. This effect is very different from other
  solvent effects which primarily influence the melting transition
  rather than the cooperativity.
\end{abstract}

\newpage

Proteins and nucleic acids (DNA and RNA) are biopolymers which are
involved in almost every biological process. These molecules are
characterized by special hierarchical structures \cite{branden_tooze}:
Each chain consists of a unique sequence of bases (DNA) or amino acids
(proteins), called the {\em primary structure}\/, which determines its
specific biological function. Short-range interactions between nearby
monomers along the chain can lead to the formation of a
three-dimensional structure, usually a helix or a planar sheet, which
is called the {\em secondary structure} of the chain. The secondary
structure is in large measure determined by the primary structure.
However, it also depends on the environmental conditions that the
polymer experiences like the temperature or the composition of the
solvent in which the chain is embedded. It is well known, for
instance, that upon increasing the temperature, the helix structure
``melts'' into a random coil structure \cite{hctransition}. Another
example is the folding-unfolding transition taking place when proteins
are stretched by external forces \cite{fold}. In this paper we study
the effects of the solvent composition on the helix-coil transition
with the emphasis on a particular effect occurring when
hydrogen-bonding agents are added to the solvent. We present a simple
model which quantifies this effect, and compare it to other solvent
effects, as well as to the effect that temperature or externally
applied force have on the transition. While we will mainly deal with
helix-coil transition occurring in polypeptides, we believe that the
model may also apply to the transition observed in DNA experiments,
just as the temperature-induced transition in both systems is studied
by means of the same model.

Polypeptides are chains of amino acids monomers linked to each other
by covalent peptide bonds. Each monomer can exist either in a
$\alpha$-helical ($h$) or a coil ($c$) state. In a helical domain
hydrogen bonding occurs along the polypeptide backbone between
monomers separated by approximately 4 units. These hydrogen bonds make
the $h$ state energetically more favorable than the $c$ state.  The
$c$ state is entropically favorable because of the rotational degrees
of freedom of the amino acids. The helical monomers at the boundaries
between $h$ and $c$ domains represent a special case: These monomers
lose their configurational entropy with no energy gain due to the
formation of hydrogen bonds. The thermodynamic behavior of the chain
can be described using the Zimm-Bragg model \cite{zimm-bragg}, in
which the free energy associated with a particular configuration (a
sequence of $h$ and $c$ monomers) of the chain is given by
\begin{equation}
F^{\rm conf}=N_h\Delta F+2N_d\Delta W.
\label{fconf}
\end{equation}
In the above equation $N_h$ denotes the number of monomers in the $h$
state, $N_d$ is the number of helical domains, $\Delta F$ is the
excess free energy per monomer in the helical state, and $\Delta W$ is
the cost in free energy to create a boundary separating $h$ and $c$
domains. The partition function is readily obtained using the transfer
matrix approach which is also applied to solve the one-dimensional
Ising model \cite{goldenfeld}. In the Zimm-Bragg model, the transition
matrix is given by
\begin{equation}
M=\left(
\begin{array}{ll}
M_{c\rightarrow c} & M_{c\rightarrow h} \\
M_{h\rightarrow c} & M_{h\rightarrow h} 
\end{array}
\right)=\left(
\begin{array}{cc}
1 & \sqrt{\sigma}s \\
\sqrt{\sigma} & s
\end{array}
\right),
\label{matrix1}
\end{equation}
where 
\begin{equation}
s=\exp(-\Delta F/kT),
\end{equation}
and 
\begin{equation}
\sigma=\exp(-2\Delta W/kT), 
\label{sigmapar}
\end{equation}
are the so-called Zimm-Bragg parameters, $k$ is the Boltzmann
constant, and $T$ is the temperature. At low temperatures $\Delta F<0$
($s>1$) since the energy gain in having an amino acid at a $h$ state
overcomes the entropy loss compared to the $c$ state. Therefore in the
low $T$ regime most of the monomers are in the $h$ state. At high
temperatures $s<1$, and the chain is mostly in the $c$ state. The
helix-coil melting transition occurs when $s\simeq 1$.  In contrast to
the significant variation of $s$ with the temperature, the parameter
$\sigma$ (also called the {\em cooperativity parameter}\/) has only
weak dependence on $T$. For biopolymers $\sigma$ is typically very
small in the range $10^{-3}-10^{-4}$. The smallness of $\sigma$
reflects the large free energy penalty in creating a helix-coil
interface, and is the reason for the sharpness of the helix-coil
transition (which, due to the one-dimensional nature of the system, is
not a true phase transition).

\begin{figure}
\begin{center}
\scalebox{.5}{\centering \includegraphics{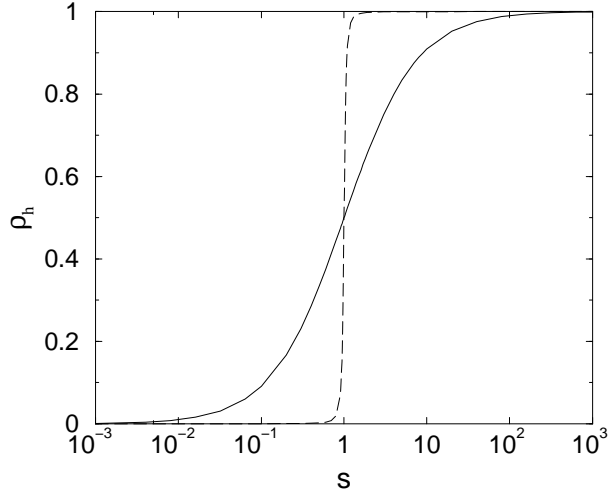}}
\end{center}
\caption{The fraction of monomers in the $h$ state, 
$\rho_h\equiv \langle N_h\rangle /N$, as a function of the Zimm-Bragg 
parameter $s$, for $\sigma=10^{-3}$ (dashed line) and $\sigma=1$ 
(solid line). The curves intersect at $s=1$, where $\rho_h=1/2$.}
\label{dohelicity}
\end{figure}

To understand the helix-coil transition in a more quantitative manner,
we write the free energy of a chain consisting of $N$ ($N\gg 1$)
monomers:
\begin{equation}
F\simeq-NkT\ln x_1,
\label{fenergy}
\end{equation}
where $x_1$ is the largest eigenvalue of the transfer matrix $M$
[Eq.(\ref{matrix1})]. The eigenvalues of $M$ are the roots of the
quadratic equation
\begin{equation}
(1-x)(s-x)-s\sigma=0.
\label{eqdeg2}
\end{equation}
The largest root of this equation is equal to
\begin{equation}
x_1=\frac{1}{2}\left[1+s+\sqrt{(1-s)^2+4s\sigma}\right].
\label{root1}
\end{equation}
The average number $\langle N_h\rangle$ of monomers in the $h$ state
is derived by differentiating the free energy [Eqs.(\ref{fenergy}) and
(\ref{root1})] with respect to $\Delta F=-kT\ln s$. We obtain
\begin{eqnarray}
\langle N_h\rangle =\frac{\partial F}{\partial (\Delta F)}=
N\left[\frac{1}{2}+\frac{s-1}{2\sqrt{(s-1)^2+4s\sigma}}\right].
\nonumber
\end{eqnarray}
The dependence of $\rho_h\equiv \langle N_h\rangle /N$, the fraction 
of monomers in the $h$ state, on $s$ is plotted in 
Fig.~\ref{dohelicity},
for $\sigma=10^{-3}$ and $\sigma=1$ (the non-cooperative case). We
observe that $\rho_h\simeq0$ for $s\ll1$, while $\rho_h\simeq1$ for
$s\gg1$.  The crossover between these two regimes occur at $s\simeq 1$
(at $s=1$, the fraction of helical monomers is exactly $1/2$). The
transition becomes sharper with decreasing $\sigma$. For small values
of $\sigma$ the width of the transition regime scales as $\Delta s
\sim \sigma^{1/2}$.

When dealing with the possible influence of the solvent composition on
the transition we should account for two types of effects. Effects of
the first type decrease the relative stability of the $h$ state over
the $c$ state. Effects of the second type reduce the helix-coil
interfacial free energy. 

Within the first group, we include changes in the $p$H or the ionic
strength of the solution that destabilize the $h$ state
\cite{soleffects1}. One can incorporate these changes in the
Zimm-Bragg model by a proper redefinition of the parameter $s$, which
now becomes a function of both the temperature and the solution
conditions. As a result of the change in the solvent conditions, the
transition (melting) temperature shifts, and it is now defined by
$s^*(T_{\rm m})=1$ with $s^*$ being the solvent-dependent parameter.

It is interesting to note that the force-induced transition, observed
in recent stretching experiments of DNA \cite{forceexperiments}, has
been also explained by the Zimm-Bragg theory with a rescaled (in this
case, a force-dependent) parameter $s$ \cite{forcetheories}. The
effect of externally applied force on the transition is usually
studied in the fixed-force ($f$) ensemble. On a mean-field level
(which turns out to provide a rather good description of the
experimentally measured elastic behavior), the elastic free energy is
given by
\begin{equation}
F^{\rm el}(f)=N_hE^{\rm el}_h(f)+(N-N_h)E^{\rm el}_c(f), 
\label{fel}
\end{equation}
where $E^{\rm el}_h(f)$ and $E^{\rm el}_c(f)$ denote, respectively,
the elastic free energy per monomer in the pure $h$ and $c$ states
(i.e., when all the monomers are in the same state). The elastic
behavior of the pure phases is usually described by one of the generic
polymer models like the freely-jointed or worm-like chain models
\cite{doi_edwards}.  Within these models, the elastic energy per
monomer is given by
\begin{eqnarray}
E^{\rm el}(f)=-\frac{1}{N}\int_0^f \!\!R(f')\,df',
\nonumber
\end{eqnarray}
where $R(f)$ is the (mean) end-to-end separation of the chain. The
transfer matrix corresponding to the sum of configurational
(\ref{fconf}) and elastic (\ref{fel}) free energies is
\begin{eqnarray}
M=\left(
\begin{array}{ll}
M_{c\rightarrow c} & M_{c\rightarrow h} \\
M_{h\rightarrow c} & M_{h\rightarrow h} 
\end{array}
\right)=e^{-E^{\rm el}_c(f)/kT}\left(
\begin{array}{cc}
1 & \sqrt{\sigma}s\, e^{[E^{\rm el}_c(f)-E^{\rm el}_h(f)]/kT} \\
\sqrt{\sigma} & s\, e^{[E^{\rm el}_c(f)-E^{\rm el}_h(f)]/kT}
\end{array}
\right),
\nonumber
\end{eqnarray}
and by comparison with the Zimm-Bragg transfer matrix (\ref{matrix1})
we readily conclude that $s^{*}(f)=s\,e^{[E^{\rm el}_c(f)-E^{\rm
    el}_h(f)]/kT}$.
 
Let us consider the change in the behavior of polypeptide chains that
occurs when the hydrogen-bonding ability of the solvent is improved.
One may consider this as another example of a solvent effect of the
first type. In alcohol environment, for instance, the hydroxide groups
tend to form hydrogen bonds with the coil monomers, thus increasing
the relative stability of the latter compared to the helical monomers.
This effect is enhanced if the hydrogen-bonding ability of the solvent
with the $c$ monomers is improved, e.g., if water is added to the
solution \cite{water}. However, one may consider the following
different scenario which serves as an example of a solvent effect of
the second type. Suppose the polypeptide chain is dissolved in an
aqueous solution, and most of the coil monomers are involved in
hydrogen bonds with the water molecules. Let us assume now that the
hydrogen-bonding ability of the solvent is improved by adding
molecules which prefer to bind to the terminal monomers of the helical
domains. The high affinity of the new hydrogen-bonding agents (which
we shall call ``impurities'') for the terminal $h$ monomers leads to
the formation of new hydrogen bonds at the interfaces between $h$ and
$c$ domains. To introduce the effect of these additional hydrogen
bonds, we need to add a term to the configurational free energy
[Eq.(\ref{fconf})]
\begin{equation}
F^{\rm conf}=N_h\Delta F+2N_d\Delta W+\mu N_{\rm i},
\label{fconf2}
\end{equation}
where $\mu$ is the chemical potential per impurity molecule attached
to the chain at the interface between $h$ and $c$ domains, while
$N_{\rm i}$ ($N_{\rm i}\leq 2N_d$) denotes the number of such impurities
\cite{remark}.  It is useful to write
\begin{equation}
\mu=\Delta\tilde{W}-\Delta W-kT\ln\varphi,
\label{chemicalpot}
\end{equation}
where $\Delta\tilde{W}$ is the free energy per ``impure'' interface
with an additional hydrogen bond (replacing $\Delta W$, the free
energy of a ``pure'' interface), while $\varphi\equiv N\xi^3/V$ is the
concentration of the {\em free}\/ impurity molecules in the solvent
(with $\xi$, the ``thermal wavelength''). The last term on the r.h.s
of Eq.(\ref{chemicalpot}) represents the loss of mixing entropy of the
impurity molecules connected to the chain compared to the free
impurities (assuming that $\varphi$ is sufficiently low). All the
other contributions to the free energy of the impure interfaces, for
instance the binding energy between the impurities and the chain, are
included in $\Delta\tilde{W}$.

In order to find the free energy corresponding to Eq.(\ref{fconf2}) we
define two new states $h^*$ and $c^*$ representing, respectively, the
terminal (``C-end'') monomer of a helical or a coil domain followed by
an impure interface. We can then apply the transfer matrix approach,
where in the present case the transfer matrix is the $4\times 4$
matrix
\begin{equation}
M=\left(
\begin{array}{llll}
M_{c\rightarrow c} & M_{c\rightarrow h} & M_{c\rightarrow c^*} & 
M_{c\rightarrow h*} \\
M_{h\rightarrow c} & M_{h\rightarrow h} & M_{h\rightarrow c^*} &
M_{h\rightarrow h^*} \\
M_{c^*\rightarrow c} & M_{c^*\rightarrow h} & M_{c^*\rightarrow c^*} 
& M_{c^*\rightarrow h^*} \\
M_{h^*\rightarrow c} & M_{h^*\rightarrow h} & M_{h^*\rightarrow c^*} 
& M_{h^*\rightarrow h^*} 
\end{array}
\right)=\left(
\begin{array}{cccc}
1 & \sqrt{\sigma}s & 1 & \sqrt{\sigma}s \\
\sqrt{\sigma} & s & \sqrt{\sigma} & s \\
0 & \sqrt{\tilde{\sigma}}s & 0 & \sqrt{\tilde{\sigma}}s \\
\sqrt{\tilde{\sigma}} & 0 & \sqrt{\tilde{\sigma}} & 0 
\end{array}
\right),
\label{matrix2}
\end{equation}
with 
\begin{eqnarray}
\tilde{\sigma}=\exp\left[-2\left(\Delta\tilde{W}-kT\ln\varphi\right)
/kT\right]=\exp\left(-2\Delta\tilde{W}/kT\right)\varphi^2.
\nonumber
\end{eqnarray}
The eigenvalues of the transfer matrix (\ref{matrix2}) are the roots
of the polynomial equation
\begin{equation}
x^2\left[(1-x)(s-x)-s(\sigma+2\sqrt{\sigma\tilde{\sigma}}
+\tilde{\sigma})\right]=0.
\label{eqdeg4}
\end{equation}
Although this is a fourth order equation, the largest root (which is
relevant to the thermodynamic behavior) comes from the quadratic
equation defined in brackets in Eq.(\ref{eqdeg4})
\begin{eqnarray}
(1-x)(s-x)-s(\sigma+2\sqrt{\sigma\tilde{\sigma}}+\tilde{\sigma})=0.
\nonumber
\end{eqnarray}
This equation is similar to the quadratic equation of the Zimm-Bragg
model [Eq.(\ref{eqdeg2})], with the cooperativity parameter $\sigma$
replaced by
\begin{equation}
\sqrt{\sigma^{*}}=\sqrt{\sigma}+\sqrt{\tilde{\sigma}}=
\sqrt{\sigma}+\exp\left(-\Delta\tilde{W}/kT\right)
\varphi.
\label{last}
\end{equation}
Our model, therefore, yields a description of the helix-coil
transition similar to the Zimm-Bragg model, but with a larger
cooperativity parameter. We find that $\sigma^{*}>\sigma$ even if the
chemical potential $\mu$ in Eqs.(\ref{fconf2}) and (\ref{chemicalpot})
is positive, i.e., if the binding of the impurities to the chain
increases the free energy of the system. This result is due to the
simple fact that some of the helix-coil interfaces will have
impurities even when $\mu>0$ and, consequently, the free energy cost
per interface will decrease (compared to the case when we had no
impurities at all, $\varphi=0$). In the ``neutral'' case $\mu=0$, for
instance, half of the hydrogen-bonding impurities will be attached to
the chain, and from Eqs.(\ref{sigmapar}), (\ref{chemicalpot}), and
(\ref{last}) we find that $\sigma^{*}=4\sigma$. We thus conclude that
the helix-coil transition becomes broader in the presence of the
hydrogen-bonding impurities, but quite unusually, the melting
temperature (which is associated with the parameter $s$) is not
predicted to change. This effect is, therefore, markedly different
from the other (``first type'') solvent effects, as well as from the
effect of externally applied force on the transition.
  
To summarize, we propose a simple description for the complicated
dependence of the helix-coil transition on the solvent character. We
suggest that solvent effects can be characterized by their impact on
the Zimm-Bragg parameters. Accordingly, we can broadly classify them
into two types: Solvent effects of the first type are associated with
a shift in the position of the helix-coil transition, and the
rescaling of the parameter $s$. Solvent effects of the second type
reduce the cooperativity of the transition, i.e., {\em increase} the
parameter $\sigma$. The amount by which $\sigma$ grows depends on the
binding free energy of the impurities, and is proportional to their
concentration in the solution. This, however, applies only to low
concentrations.  At larger densities it is necessary to include higher
virial coefficients in the mixing-entropy term [Eq.\ref{chemicalpot})]
to describe the interactions between the impurities.

We thank the referee for his very useful remarks, and M.~C.~Williams
for bringing our attention to his experimental work on this subject.
We also acknowledge discussions with K. Plaxco, J. Israelachvili and
T. Deming. The work was supported by the MRL Program of the National
Science Foundation under Award No.~DMR00-80034.



\begin{thebibliography}{99}
\bibitem{remark} In {\protect Eq.(\ref{fconf2})} we assume that all
  the new hydrogen bonds are formed at the interfaces between helix
  and coil domains. This assumption applies to the case when the free
  energy gain in creating such a new bond is much larger than the
  free energy gain in replacing an existing bond between a coil
  monomer and a water molecule, with a bond between a coil monomer and
  impurity.
\end{thebibliography}
\end{document}